\documentclass[aps,prl,twocolumn]{revtex4}
\usepackage{amssymb}

\usepackage{graphicx}

\begin{document}

\title{Orbitally quantized density-wave states perturbed from equilibrium }
\author{N.~Harrison, R.~D.~McDonald, and J. Singleton}
\affiliation{National High Magnetic Field Laboratory, LANL, Los Alamos, NM 87545}
\date{\today}

\begin{abstract}
We consider the effect that a change in the magnetic induction ${\bf B}$ has in causing an orbitally quantized field-induced spin- or charge density wave (FISDW or FICDW) state to depart from thermodynamic equilibrium. The competition between elastic forces of the density wave (DW) and pinning leads to the realization of a critical state that is in many ways analogous to that realized within the vortex state of type II superconductors. Such a critical state has been verified experimentally in charge-transfer salts of the composition $\alpha$-(BEDT-TTF)$_2M$Hg(SCN)$_4$, but should be a generic property of all orbitally quantized DW phases. The metastable state consists of a balance between the DW pinning force and the Lorentz force on extended currents associated with drifting cyclotron orbits, resulting in the establishment of persistent currents throughout the bulk and to the possibly of
a three-dimensional `chiral metal' that extends deep into the interior of a crystal. 
\end{abstract}

\pacs{PACS numbers:
..............................}
\maketitle

\section{introduction}
There is growing interest in broken-translational-symmetry phases that incorporate orbital quantization~\cite{brandt1,tsui1,ribault1,williams1,hill1,harrison1,hoffman1,andres1,harrison2,graf1}. The interplay between orbital and periodic charge, spin or current degrees of freedom introduces additional constraints that can lead to new types of quantum order with radically different physical properties. Organic charge-transfer salts based on the TMTSF molecule provide us with a particularly vivid example. Orbital quantization constrains the allowed values of the spin-density modulation vector ${\bf Q}_\nu$ for a given Landau level filling factor $\nu$, leading to discrete field-induced spin-density wave (FISDW) phases with different quantized Hall conductances~\cite{ribault1,hannahs1,chaikin1}. One consequence of orbital quantization of the energy spectrum into levels (or subbands), separated in energy by $\hbar\omega_{\rm c}=\hbar e|{\bf B}|/m$, that is common to all these systems~~\cite{brandt1,tsui1,ribault1,williams1,hill1,harrison1,hoffman1,andres1,harrison2,graf1}, is the dependence of the equilibrium modulation vector $\bar{\bf Q}({\bf B})$ (or the period $\lambda=2\pi/|\bar{\bf Q}({\bf B})|$) on the magnetic induction ${\bf B}$. Standard theoretical models are concerned with FISDW or field-induced charge-density wave (FICDW) phases that are in thermal equilibrium~\cite{ribault1,andres1,chaikin1}, with ${\bar{\bf Q}}({\bf B},{\bf r})$ being uniform over all spatial coordinates ${\bf r}$. 

The present review considers the effect that density wave (DW) pinning has on changing the physical properties of orbitally quantized DW states~\cite{gruner1}. Pinning inhibits the establishment of thermodynamic equilibrium as ${\bar{\bf Q}}({\bf B},{\bf r})$ changes with ${\bf B}$, leading to the induction of non-equilibrium (metastable) states in response to a change in the magnetic field ${\bf H}$. This may simply be a direct change in the externally applied magnetic field ${\bf H}={\bf B}/\mu_0-{\bf M}$ or the field gradient associated with the application of an electrical transport current ${\bf j}=\nabla\times{\bf H}$. Competition between pinning and the elastic DW restoring force (due to its perturbation from equilibrium) results in a critical state analogous to that encountered in the vortex state of type II superconductors~\cite{senoussi1}. The critical region consists of a current associated with the drift of cyclotron orbits orthogonal to the pinning force that has an equivalent role to a pinned supercurrent. The pinning potential also modifies the Eigenvalues of the quasi-two dimensional electron (or hole) gas, giving rise to the possibility of a type of `chiral metal' that extends deep into the bulk, rather than being restricted to the surface~\cite{balents1}. The finite spatial extent of the critical regions have the potential to radically transform the magnetotransport properties of the system as a whole.  This can be an important  
factor in the realization of large Hall angles in bulk FISDW and FICDW systems, greatly enhanced conductances and persistent current phenomena~\cite{harrison2,hannahs1,chaikin1,honold1}. We consider the specific cases of metastable FISDW and FICDW states in $\alpha$-(BEDT-TTF)$_2M$Hg(SCN)$_4$ and (TMTSF)$_2X$. 
\section{Critical state}
\begin{figure}
\includegraphics[width=0.4\textwidth]{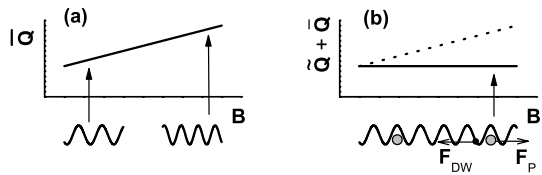}
\caption{(a) Schemtic example of an equilibrium modulation vector $\bar{\bf Q}$ that increases with ${\bf B}$, with a sketch showing how the period becomes shorter. (b) A plot showing how a pinning force ${\bf F}_{\rm p}$ prevents equilibrium (dotted line) from being achieved, leading to a non-equilibrium contribution $\tilde{\bf Q}$ (solid line).}
\label{diagram}
\end{figure}

To understand the physics of pinning in the context of orbitally quantized DW states, it is instructive to refer to the simple model depicted in Fig.~\ref{diagram}. We consider a linear relationship between $\bar{\bf Q}$ and ${\bf B}$, such that an increase in ${\bf B}$ causes the equilibrium period $\lambda$ of the spin- or charge-density wave modulation to shorten. Since real samples are of finite spatial extent, maintenance of equilibrium requires the density modulation to undergo translation with respect to ${\bf r}$ in accordance with the continuity equation
\begin{equation}\label{continuity}
\frac{\partial\rho_\lambda}{\partial t}+\nabla\cdot{\bf j}_\lambda=0.
\end{equation}
Here, $\rho_\lambda$ represents the physical quantity (e.g. charge, spin or current) affected by the modulation and ${\bf j}_\lambda$ represents the current associated with its translational motion. Pinning inhibits maintenance of equilibrium, leading to non-equilibrium contribution $\tilde{\bf Q}$ such that the total ${\bf Q}=\tilde{\bf Q}+\bar{\bf Q}$ remains initially unchanged as depicted in Fig.~\ref{diagram}b. The stored energy associated with the compression (or tension) of the density modulation is therefore analogous to that of a spring, with pinning sites then opposing the restoring force ${\bf F}_{\rm DW}$ by providing an equal and opposite pinning force ${\bf F}_{\rm p}$.

\begin{figure}
\centering \includegraphics*[scale=0.9,angle=0]{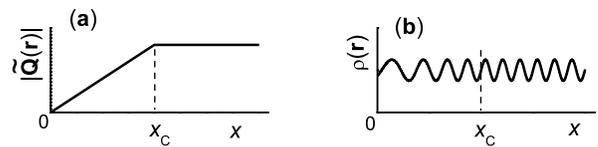}
\caption{A notional representation of a `critical state' region between ${\bf r}=$~0 and ${\bf r}=[x_{\rm c},0,0]$ depicting (a) the gradient in $\tilde{\bf Q}({\bf r})$ and (b) the gradient in the modulation density.}
\label{critical}
\end{figure}

The physics of exactly this situation has been extensively studied in the vortex state of type II superconductors~\cite{senoussi1}; in this case involving the spatial modulation of the supercurrent density of a flux-line lattice. The field-induced compression (or tension) of the modulation with respect to equilibrium in Fig.~\ref{diagram}b leads to a build up of non-equilibrium stored energy 
\begin{equation}\label{vortexenergy}
\tilde{\Phi}=\frac{\mu_0(\Delta {\bf H})^2}{2}\propto(\Delta\tilde{\bf Q})^2.
\end{equation}
The surface of a sample then plays a pivotal role in determining how the system responds to a change $\Delta{\bf H}$ in the external magnetic field ${\bf H}$. Since pinning cannot support the infinite Lorentz force per vortex ${\bf F}_{\rm DW}\propto\nabla\partial\tilde{\Phi}/\partial\rho_\lambda$ that would otherwise result were $\tilde{\Phi}$ to remain finite up to the sample surface, the vortex lattice undergoes translational motion so as to ensure that $\tilde{\Phi}=0$ at the surface. This initiates the formation of a critical state, whereupon the Lorentz force ${\bf F}_{\rm DW}$ is balanced by the maximum pinning force ${\bf F}_{\rm p,max}$ that the sample can sustain. The critical region then propagates progressively further into the interior of the sample as the external magnetic field ${\bf H}$ is changed. The point $x=x_{\rm c}$ in Fig.~\ref{critical} represents the furthest extent of the critical region into the sample for a given change $\Delta{\bf H}$ in ${\bf H}$. Within the `critical' region (i.e. 0~$<x<x_{\rm c}$) there exists a gradient in $\tilde{\bf Q}({\bf r})$ and $\lambda$, while  beyond the critical region (i.e. $x>x_{\rm c}$) $\tilde{\bf Q}({\bf r})$ remains approximately uniform.

\begin{figure}
\centering \includegraphics*[scale=0.45,angle=0]{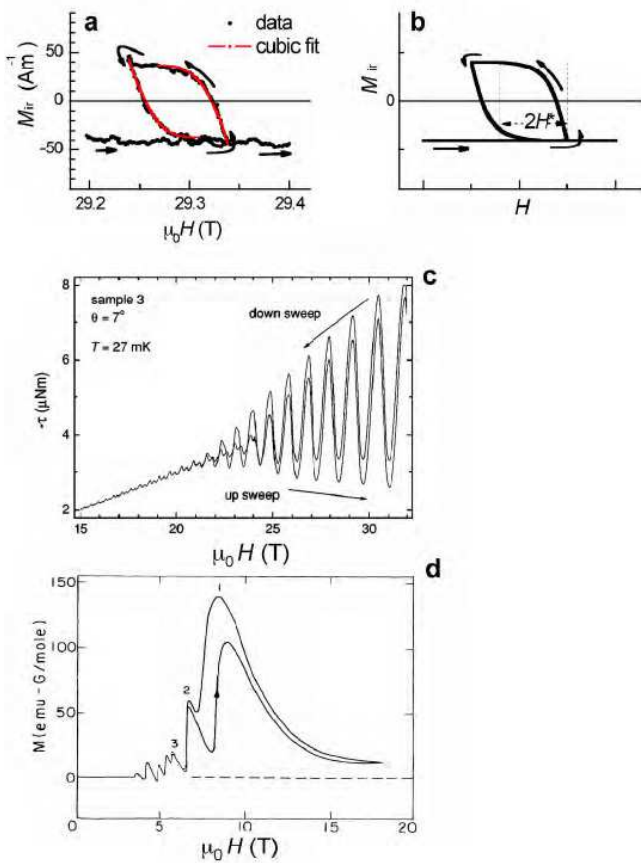}
\caption{Examples of the non-equilibrium magnetization of field-induced density-wave measured by way of magnetic torque. (a) shows a hysteresis loop obtained on cycling the magnetic field in $\alpha$-(BEDT-TTF)$_2$KHg(SCN)$_2$~\cite{harrison2}, together with a cubic fit (red). This plot shows only the irreversible contribution $M_{\rm irr}$ (the reversible pure de~Haas-van~Alphen part has been subtracted). (b) Shows a simulation of a hysteresis loop for a hypothetical cylindrical sample, resulting in the cubic lineshape where $H^\ast$ is the coercion field. Other geometries will yield a cubic lineshape to leading order. (c) shows the consequences of such hysteresis over an extended range of magnetic field leading to different up and down magnetic field sweeps, measured on a different sample~\cite{harrison1}. (d) shows qualitatively similar hysteresis over an extended range of magnetic field in (TMTSF)$_2$ClO$_4$ in which FISDW states are realized at ambient pressure~\cite{naughton1}.}
\label{hysteresisexamples}
\end{figure}

Since the build up of non-equilibrium stored energy $\Phi(\tilde{\bf Q})$, orbital quantization and pinning are common to flux-line lattices and field-induced DW states, the response of both types of system to a change in magnetic field exhibits a high degree of similarity~\cite{harrison2,harrison4}. The primary difference in the case of DW systems is that the quasiparticles are non-superconducting (i.e. they have a finite relaxation time) causing the irreversible diamagnetic susceptibility $\tilde{\chi}=\partial M/\partial H$ to depart significantly from the ideal diamagnetic value of $-1$. A critical state nevertheless develops, yielding magnetic hysteresis that can be observed experimentally within orbitally quantized DW states. The existence of such a critical state has been directly demonstrated in charge-transfer salts of the composition $\alpha$-(BEDT-TTF)$_2M$Hg(SCN)$_2$ (where $M=$~K and Rb)~\cite{harrison1,harrison3,harrison2,harrison4}. Figure~\ref{hysteresisexamples}a-c show examples in $\alpha$-(BEDT-TTF)$_2$KHg(SCN)$_2$.
A qualitatively similar magnetic hysteresis occurs deep within the $\nu=$~1 FISDW state of (TMTSF)$_2$ClO$_4$~\cite{naughton1} (shown in Fig.~\ref{hysteresisexamples}d), although the formation of a critical state remains to be studied. 

\section{Model for non-equilibrium field-induced density-wave states}
To model the critical state in orbitally quantized DW states, it is convenient to consider the surface of a three-dimensional crystal normal to the unit vector $\hat{\bf x}$, as represented graphically in Fig.~\ref{critical}.
This enables us to consider a situation in which the components of the electric field  ${\bf E}=[E_x(x),E_y(x),0]$, current ${\bf j}=[0,j_y(x),0]$ and magnetic flux density ${\bf B}=[0,0,B_z(x)]$ vary 
only with respect to $x$ in the continuity limit (i.e. $E_x/(\partial E_x/\partial x)\gg\lambda$). We consider DW formation and Landau quantization to take place within layers 
orthogonal to the unit vector $\hat{\bf z}$. The total magnetic flux density $B_z=B_0+\Delta B_z$ parallel to $\hat{\bf z}$ is composed of a uniform component $B_0=\mu_0H_0$ outside the sample, and a non-uniform perturbation $\Delta B_z(x)$ inside the sample. The gradient $-\partial\Delta B_z/\mu_0\partial x$ can be considered to be the sum $j_y=j_{y,{\rm f}}+j_{y,{\rm m}}$ of free  $j_{y,{\rm f}}=-\partial\Delta H_z/\partial x$ and pinned (or magnetic) $j_{y,{\rm m}}=-\partial\Delta M_z/\partial x$ currents, such that $\Delta B_z=\mu_0(\Delta H_z+\Delta M_z)$. Free currents $j_{y,{\rm f}}$ are those that contribute directly to the electrical transport in response to ${\bf E}$ while magnetic currents $j_{y,{\rm m}}$ are those resulting from the drift of extended states orthogonal to the pinning force ${\bf F}_{\rm p}$.

\subsection{Materials of interest}
\begin{figure}
\centering \includegraphics*[scale=0.8,angle=0]{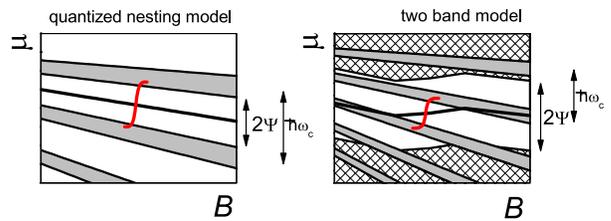}
\caption{A schematic of part of the energy level structure over a short range of $\mu$ and ${\bf B}$ for the `quantized nesting' and `two band' models. Grey shading represents Landau levels or Landau subbands while the hatched area represents the ungapped density of states of a non-orbitally quantized SDW or CDW in the two band model. The thick black and red lines represent loci of $\Delta\bar{\mu}(B_z)$ and $\Delta\bar{\mu}(B_z)+\Delta\tilde{\mu}(B_z)$ respectively.}
\label{model}
\end{figure}

We consider two qualitatively different models of orbitally quantized DW states that apply to bulk crystalline materials, that have already been shown to be readily accessible to inductive experiments. The first Landau level spectrum depicted graphically in the left panel of Fig.\ref{model} corresponds to the `quantized nesting model'~\cite{chaikin1} that is generally representative of FISDW states in (TMTSF)$_2X$ salts (where $X=$~PF$_6$, AsF$_6$ or ClO$_4$).  A variant of this model may also apply to $\alpha$-(BEDT-TTF)$_2M$Hg(SCN)$_4$~\cite{andres1} (where $M=$~K, Tl or Rb) for ${\bf B}$ applied orthogonal to the $b$-axis, or to (Per)$_2$Pt(mnt)$_2$~\cite{graf1}, provided there exists sufficient overlap of electronic orbitals between Perylene chains. The second energy level spectrum depicted graphically in Fig. \ref{model} corresponds to a two band model in which the orbital quantization of a 2D hole pocket and DW formation occur on separate bands as in $\alpha$-(BEDT-TTF)$_2M$Hg(SCN)$_4$~\cite{harrison2,harrison4} (where $M=$~K, Tl or Rb) but that nevertheless share the same chemical potential $\mu$. In both cases, we assume that orbital quantization occurs only in layers orthogonal to $\hat{\bf z}$ and that the two spin states are degenerate. The latter occurs naturally as a consequence of SDW formation in strong magnetic fields when the spins lie orthogonal to ${\bf B}$~\cite{chaikin1} or may also occur in a CDW system when the product $mg$ (where $g$ is the {\it g}-factor) is an integer number of free electron masses $m_{\rm e}$, as is approximately the case in $\alpha$-(BEDT-TTF)$_2$KHg(SCN)$_4$~\cite{harrison2,harrison4}.  
 
In spite of the differences in electronic structure, the non-equilibrium behaviors of these model systems are equivalent. ${\bf B}$ equally influences both systems through a change in the Landau level degeneracy $D=m\omega_{\rm c}/\pi\hbar c$, where the cyclotron frequency is given by~\cite{harrison2,macdonald1}
\begin{equation}\label{cyclotron}
\omega_{\rm c}=\frac{qB_z}{m}\bigg(1+\frac{qm\nabla\cdot{\bf E}}{2e^2B_z^2}\bigg)^{1/2}.
\end{equation}
Here, $q$ is the charge of the quasiparticles (positive for holes and negative for electrons) while `$c$' is the interlayer spacing. The second term inside the bracket accounts for a possible divergence $\nabla\cdot{\bf E}$ in the electrical field due to a small non-equilibrium charge density within the sample. This might occur were the diagonal components of the resistivity tensor $\rho_{xx}$, $\rho_{yy}$ and $\rho_{zz}$ to vanish, should the transport become ballistic under certain conditions (see Section 4.3). The derivation of this term has been covered in detail in the literature~\cite{harrison2,macdonald1}; hence here we merely quote the result. The total degeneracy can be expressed as
\begin{equation}\label{degeneracy}
D_0+\Delta D=\frac{q}{\pi\hbar c}\bigg(B_0+\Delta B_z+\frac{m}{2qB_0}\frac{\partial E_x}{\partial x}\bigg)
\end{equation}
for the geometry considered in Fig.~\ref{critical}, where $\Delta D$ is then its perturbation from equilibrium. The mean density of states $\bar{g}_{\rm 2D}=D/\hbar\omega_{\rm c}=m/\pi\hbar^2c$  (averaged over energy) remains independent of $B_z$. 

The response of the system to $\Delta B_z$ depends on changes in the relative position of $\mu$ between the Landau levels in Fig.~\ref{model}. This can be tracked by considering the fact that the total perturbed charge density of the 2D Landau levels ($\Delta\rho_{\rm 2D}$) and of the DW ($\Delta\rho_{\rm DW}$) must obey Poisson's equation,
\begin{equation}\label{neutrality}
\Delta\rho_{\rm 2D}+\Delta\rho_{\rm DW}=-\varepsilon\nabla\cdot{\bf E},
\end{equation}
hence preserving charge neutrality when $\nabla\cdot{\bf E}=$~0 (in the continuity limit). The most extreme situation occurs when the chemical potential $\mu$ resides between Landau levels, as depicted in Fig.~\ref{model}. The perturbed charge density contribution due to the Landau levels in Fig.~\ref{model}, for both DW models, is then given by
\begin{equation}\label{2Dcharge}
\Delta\rho_{\rm 2D}=\nu q\Delta D-\nu\beta q\Delta D-\beta q\bar{g}_{\rm 2D}\Delta\mu,
\end{equation}
where the first term on the right-hand side represents the sum over all $\nu$ occupied Landau levels which shift in response to $\Delta B_z$. The second term on the right-hand side is a correction (constant $\mu$) for the existence of a finite density of states $\beta\bar{g}_{\rm 2D}$ between the Landau levels. If the Landau levels are broadened by a finite quantum lifetime $\tau$, for example, then $\beta\approx 4/\pi\omega_{\rm c}\tau\ll$~1. The third term on the right-hand side accounts for the change $\Delta\mu$ in $\mu$. The latter is related to $\Delta{\bf Q}$ by
\begin{equation}\label{chemicalpotential}
\Delta\mu=-{\rm sign}(q_{\rm DW})\frac{\hbar}{2}\bar{\bf v}_{\rm F}\cdot\Delta{\bf Q},
\end{equation}
where $\bar{\bf v}_{\rm F}$ is the mean Fermi velocity of the band on which the DW forms and ${\rm sign}(q_{\rm DW})$ represents the sign of its carriers. We have $q_{\rm DW}\equiv q$ in the case of the quantized nesting model in Fig.~\ref{model}, whereas $q_{\rm DW}=-q$ in the two band model. The equilibrium and non-equilibrium evolution of the DW is considerably simplified by using the variable $\Delta\mu$ instead of $\Delta{\bf Q}$. In a similar fashion to $\Delta{\bf Q}$, $\Delta\mu$ must consist of both equilibrium $\Delta\bar{\mu}$ and non-equilibrium $\Delta\tilde{\mu}$ components such that
\begin{equation}\label{mu}
\Delta\mu=\Delta\bar{\mu}+\Delta\tilde{\mu}.
\end{equation}

\subsection{Equilibrium conditions}
For the DW to achieve the equilibrium state described in standard theoretical models, the DW must adjust itself according to Equation (\ref{continuity}) so as to minimize the total free energy~\cite{harrison2} (see below). This is acccomplished when $\Delta\tilde{\mu}=$~0 in Equation (\ref{mu}), in which case $\mu$, represented by the thick black line in Fig.~\ref{model}, lies in the middle of the DW gap. This occurs when
\begin{equation}\nonumber
\Delta\rho_{\rm DW}=-\nu q_{\rm DW}\Delta D~~~{\rm \{quantized~nesting\}}
\end{equation}\begin{equation}\label{DWcharge}
\Delta\rho_{\rm DW}=q_{\rm DW}\bar{g}_{\rm DW}\Delta\bar{\mu}~~~{\rm \{two~band\}}~~~~~~~~~
\end{equation}
for the quantized nesting and the two band models respectively. On inserting each expression for $\Delta\rho_{\rm DW}$ (together with Equations (\ref{2Dcharge}) and (\ref{degeneracy})) into Equation (\ref{neutrality}), we obtain
\begin{equation}\label{equilibrium}
\frac{m}{2qB_0}\frac{\partial E_x}{\partial x}+\Delta B_z=\alpha\frac{m}{\hbar q\nu}\Delta\bar{\mu},
\end{equation}
where $\alpha\equiv-1$ and $\alpha=(\eta+\beta)/(1-\beta)\approx\eta$ for the for the quantized nesting and two band models respectively. Note that the right-hand side of Equation (\ref{neutrality}) is much smaller than the similar contribution $\propto\nabla\cdot{\bf E}$ originating from Equation (\ref{cyclotron}) and can therefore be neglected. 
 
\subsection{Pinning and non-equilibrium thermodynamics}
As discussed in the introduction, pinning inhibits maintenance of an equilibrium DW state by preventing it from moving (or sliding)~\cite{gruner1} so as to satisfy the continuity Equation (\ref{continuity}).
The DW is no longer able to respond to $B_z$ for $x>x_{\rm c}$ in Fig.~\ref{critical}, instead remaining physically static such that $\Delta\rho_{\rm DW}=$~0. Under these circumstances $\Delta\tilde{\mu}\neq$~0, with  Equation (\ref{2Dcharge}) therefore yielding
\begin{equation}\label{nonequilibrium1}
\frac{m}{2qB_0}\frac{\partial E_x}{\partial x}+\Delta B_z=\frac{\beta m}{(1-\beta)\hbar q\nu}(\Delta\tilde{\mu}+\Delta\bar{\mu}).
\end{equation}
Equation (\ref{nonequilibrium1}) describes the red line in Fig.~\ref{model}, which yields precisely the same result for both the quantized nesting and two-band models.
On substituting $\Delta\bar{\mu}$ from Equation(\ref{equilibrium}), this reduces to 
\begin{equation}\label{nonequilibrium2}
\frac{m}{2qB_0}\frac{\partial E_x}{\partial x}+\Delta B_z=\gamma\frac{\beta m}{\hbar
q\nu}\Delta\tilde{\mu},
\end{equation}
where $\gamma=1$ and $\gamma=(\eta+\beta)/\eta(1-\beta)\approx$~1 for the quantized nesting and two band models respectively. Because $\beta<<1$ in the right-hand side of Equation (\ref{nonequilibrium2}), the non-equilibrium state of the DW is {\it rapidly} established, even for  a very small $\Delta B_z$.

As with the critical state model of type II superconductors, any perturbation of the orbitally quantized DW state from equilibrium leads to the build up of stored energy. In the case of the DW, this energy is given to leading order by
\begin{equation}\label{freeenergyDW}
\bar{\Phi}_{\rm DW}+\tilde{\Phi}_{\rm DW}\approx-g_{\rm DW}\Bigg(\frac{\Psi^2}{2}-\big(\frac{\hbar v_{\rm F}}{2}\big)^2\bigg(\frac{\partial\phi}{\partial x_{\bf Q}}\bigg)^2\Bigg)
\end{equation}
where $\partial\phi/\partial x_{\bf Q}$ represents the gradient in the phase $\phi$ of the DW parallel to ${\bf Q}$~\cite{gruner1}. The first term on the right-hand-side is the equilibrium part whereas the second term is the non-euilibrium part. Through Equation (\ref{chemicalpotential}), this second term is equivalent to $\Delta\tilde{\mu}$~\cite{harrison6}. The magnetization change associated with 2D Landau level spectrum contributes as additional component to the free energy, which is given to leading order by 
\begin{equation}\label{freeenergy2D}
\tilde{\Phi}_{\rm 2D}\approx\beta g_{\rm 2D}\frac{\Delta\tilde{\mu}^2}{2}.
\end{equation}
Were $\beta=$~0, as for a pristine system with an infinite quantum lifetime, pinning of the DW would cause $\mu$ to jump discontinuously between Landau levels, with no net change in the free energy $\tilde{\Phi}_{\rm 2D}$. This is a consequence of the fact that the free energy of a 2D electron gas in the canonical ensemble retains its minimum value irrespective of the location of $\mu$ within the gap. It is the existence of a finite density of states $\approx\beta g_{\rm 2D}$ within the gap that causes $\tilde{\Phi}_{\rm 2D}$ to become dependent on $\mu$~\cite{harrison7}; hence the $\beta$ prefactor to Equation (\ref{freeenergy2D}).

On combining Equations (\ref{freeenergyDW}) and (\ref{freeenergy2D}), we obtain
\begin{equation}\label{freeenergy}
\tilde{\Phi}\approx (g_{\rm DW}+\beta g_{\rm 2D})\Delta\tilde{\mu}^2,
\end{equation}
for the total non-equilibrium stored energy. It is immediately apparent on inspection of Equation (\ref{freeenergy}) that the contribution $\propto\beta g_{\rm 2D}$ from the magnetization of the 2D Landau level spectrum is much smaller than that from the DW. This implies that the effective magnetic current originating from the gradient in the de~Haas-van~Alphen magnetization due to a gradient in $\Delta\tilde{\mu}$ caused by pinning is much smaller than that associated directly with the drift of the cyclotron orbits orthogonal to the pinning force. As will soon become apparent, it is this latter contribution that is entirely responsible for the rather unusual behavior of pinned orbitally quantized DW systems. On neglecting the smaller contribution from the 2D Landau levels, we arrive at 
\begin{equation}\label{freeenergyBE}
\tilde{\Phi}\approx g_{\rm DW}\big(\frac{\hbar\nu}{\beta\gamma}\big)^2\bigg(
\frac{1}{2B_0}\frac{\partial E_x}{\partial x}+\frac{q}{m}\Delta B_z\bigg)^2.
\end{equation}
The substitution of Equation (\ref{nonequilibrium2}) into Equation (\ref{freeenergy}) enables us to express the stored energy in terms of the perturbations $\Delta B_z$ and $\partial E_x/\partial x$. If $\rho_{xx}$, $\rho_{yy}$ and $\rho_{zz}$ remain finite, we can assume that $\partial E_x/\partial x\rightarrow$~0 in a 
slowly varying magnetic field. On taking the second derivative with respect to $\Delta B_z$, we then obtain
\begin{equation}\label{2ndderivative}
\frac{\partial\Delta\tilde{M}_z}{\partial B_z}=-\frac{\partial^2\tilde{\Phi}}{\partial B_z^2}=
-2g_{\rm DW}\bigg(\frac{\hbar\nu e}{\beta\gamma m}\bigg)^2,
\end{equation}
with the non-equilibrium susceptibility given by
\begin{equation}\label{susceptibility}
\tilde{\chi}=\frac{\partial\Delta\tilde{M}_z}{\partial H_z}=\Bigg(\bigg(\mu_0\frac{\partial\Delta\tilde{M}_z}{\partial B_z}\bigg)^{-1}-1\Bigg)^{-1}.
\end{equation}
For the purposes of making numerical estimates of $\tilde{\chi}$, it is convenient to express Equation (\ref{susceptibility}) in terms of more familar parameters
\begin{equation}\label{susceptibility2}
\tilde{\chi}\approx\bigg(\bigg(2\mu_0\eta g_{\rm 2D}\big(\frac{\pi\varepsilon_{\rm F}\mu_{\rm m}}{4}\big)^2\bigg)^{-1}-1\bigg)^{-1},
\end{equation}
where $\varepsilon_{\rm F}\approx\hbar e\nu B_0/m\approx N_{\rm 2D}/g_{\rm 2D}$ is the Fermi energy scale of the 2D pocket, $g_{\rm 2D}=m/\pi\hbar^2c$ is its density of states and $\mu_{\rm m}=e\tau/m$ is its carrier mobility. 

\begin{figure}
\centering \includegraphics*[scale=1.5,angle=0]{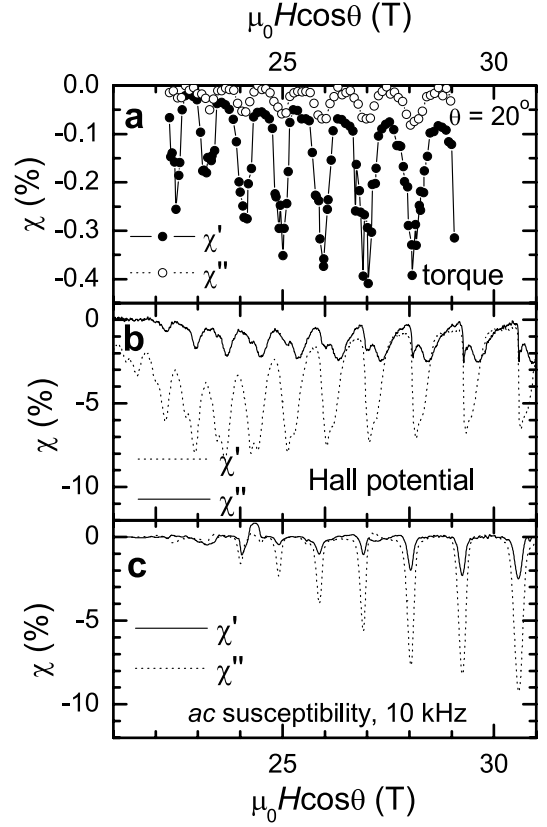}
\caption{Differential susceptibility of $\alpha$-(BEDT-TTF)$_2$KHg(SCN)$_4$ obtained using different methods. (a) From the maximum rate of change in magnetic torque with field on measuring many hysteresis loops like those in Fig.~3 at different fields and $\sim$~30~mK~\cite{harrison2}. (b) From oscillations in the Hall voltage measured between the edge and the center of the uppermost surface layer of a crystal with the excitation provided by means on an ac magnetic field of $\sim$~400~Hz superimposed on a static magnetic field at $\sim$~0.5~K. (c) Using the conventional ac susceptibility method in a static background magnetic field at $\sim$~0.5~K. In all cases, $\theta<$~20$^\circ$.}
\label{susceptibilitydata}
\end{figure}

For (TMTSF)$_2$ClO$_4$, where $\eta=$~1 (since the DW and Landau quantization involve the same band), $m\approx$~0.1~$m_{\rm e}$, $c\approx$~14~\AA, $g_{\rm 2D}\approx$~2~$\times$~10$^{45}$ m$^{3}$J$^{-1}$, $\varepsilon_{\rm F}\approx$~23~meV and $\mu_{\rm m}\approx$~1~T$^{-1}$~\cite{ishiguro1,chaikin1,lebed1}, we obtain $\tilde{\chi}\approx$~-0.04. For $\alpha$-(BEDT-TTF)$_2$KHg(SCN)$_4$, where $\eta\approx$~0.5, $m\approx$~2~$m_{\rm e}$, $c\approx$~20~\AA (usually referred to as b), $g_{\rm 2D}\approx$~3~$\times$~10$^{46}$ m$^{3}$J$^{-1}$, $\varepsilon_{\rm F}\approx$~38~meV and $\mu_{\rm m}\approx$~0.5~T$^{-1}$~\cite{harrison1}, we obtain an estimate of $\tilde{\chi}\approx$~-0.1 that is slightly larger. The latter estimate for $\tilde{\chi}$ is also roughly in agreement with estimates made from magnetic torque and ac susceptibilty measurements shown in Fig.~\ref{susceptibilitydata}~\cite{harrison2}.

\section{Magnetotransport}
\begin{figure}
\centering \includegraphics*[scale=1,angle=0]{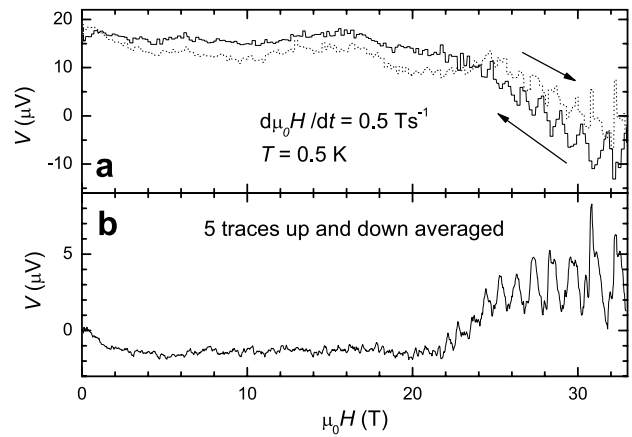}
\caption{The induced Hall voltage measured between the edge and the center of the uppermost surface layer of a crystal on sweeping the magnetic field up and down at $\sim$~0.5~K. (a) Shows raw data while (b) shows 5 up and down sweeps averaged.}
\label{bittersweep}
\end{figure}

\begin{figure}
\centering \includegraphics*[scale=0.8,angle=0]{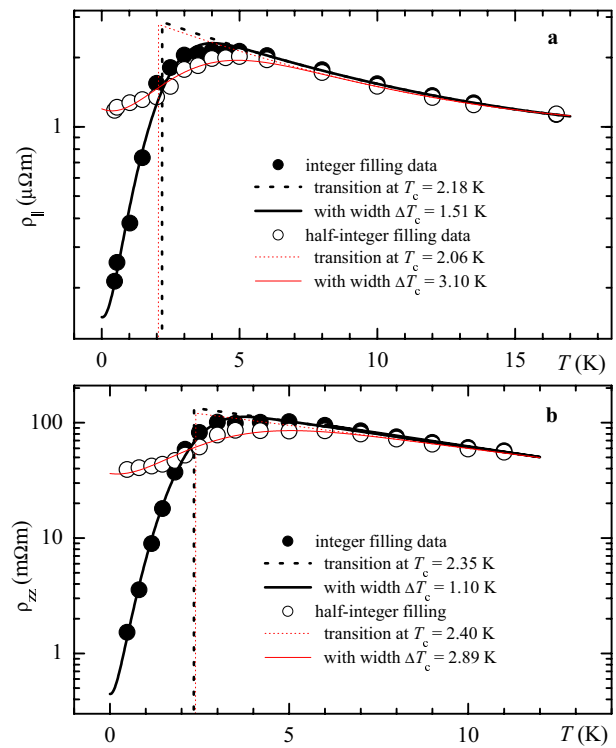}
\caption{Temperature dependent intra-layer $\rho_\|\approx\rho_{xx}$ (a) and inter-layer $\rho_{zz}$ (b) resistivity of $\alpha$-(BEDT-TTF)$_2$KHg(SCN)$_4$ at integer (filled circles) and half-integer (open circles) filling factors at $\mu_0H\sim$~30~T, together with fits of an error function at $T_{\rm c}$ of width $\Delta T_{\rm c}$ (solid lines) renormalized by the extrapolated normal state resistivity. The dotted lines are obtained on setting $\Delta T_{\rm c}=$~0 in the fitted function.}
\label{transition}
\end{figure}

The magnetotransport of orbitally quantized DW states in equilibrium has been examined throughout the literature. The consequences of non-equilibrium thermodynamic effects are more difficult to predict. The experimental magnetotransport has proven to be rather anomalous compared to ordinary organic conductors. For example, in the case of $\alpha$-(BEDT-TTF)$_2$KHg(SCN)$_4$, an induced Hall potential can be detected by slowly sweeping the field of a Bitter magnet (shown in Fig.~\ref{bittersweep}). This result can be partly interpreted in terms of an unusually large Hall ratio $\rho_{xy}/\rho_{xx}$ of $\sim$~180~\cite{harrison2}. The development of quantized Hall plateaux in $\rho_{xy}$, a large Hall ratio and reduced longitudinal resistivity $\rho_{xx}$ at low temperatures are the natural consequence of the opening of a Landau gap, as demonstrated in (TMTSF)$_2$AsF$_6$~\cite{chaikin1}. Such effects appear to be particularly severe in $\alpha$-(BEDT-TTF)$_2$KHg(SCN)$_4$, however, with the change in resistivity as a function of temperature on entry into the field-induced CDW state resembling that of a phase transition into a dissipationless conducting state (see Fig.~\ref{transition}~\cite{harrison5}).

There are several ways in which one can attempt to understand such anomalous data. One is to assume scattering to be an essentially random process, resulting in uniformly broadened Landau levels throughout the bulk, and hence a uniform $\beta$. Another is to assume that scattering occurs mostly at spatially extended defects, causing the relative position of the Landau level Eigenvalues to vary throughout the bulk. Alternatively, $\beta$ can vary. In this case $\beta$ will depend on the local position of $\mu$ with respect to the Landau gap as well as on the local concentration of impurities. Another is to consider more radical changes in the potential landscape introduced by the critical state, leading to the possibility of a bulk chiral metal. Below we consider each of these in turn.

\subsection{Uniform transport}
\begin{figure}
\centering \includegraphics*[scale=0.8,angle=0]{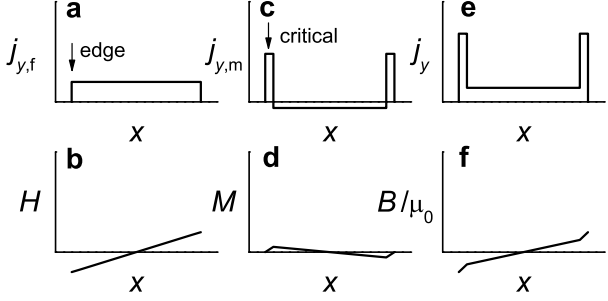}
\caption{Current and magnetic field distributions across the width of a hypothetical bar-like sample. (a) represents a uniform applied current giving rise to the variation in $H$ in (b). (c) represents the induced current due to $\tilde{\chi}$ giving rise to no net magnetization $M$ in (d) across the sample owing to the critical region at the edge. (e) and (f) show the total current and magnetic induction $B$.}
\label{currents}
\end{figure}

Even in the simplest case of a uniform sample (with a uniform $\beta$),
the simple action of applying a transport current $j_{y,{\rm f}}$ within the bulk perturbs the system from equilibrium. A uniform current density, like that depicted in Fig.~\ref{currents}a for example, gives rise to a gradient in the magnetic field $H_z$ along $\hat{\bf x}$ in Fig.~\ref{currents}b, which through Equation (\ref{nonequilibrium2}) gives rise to a gradient $\nabla\Delta\tilde{\mu}$ in $\Delta\tilde{\mu}$. This gradient exacts an additional force
\begin{equation}\label{force}
\tilde{\bf F}_{\rm DW}=-\frac{2\eta}{\gamma\beta}\nabla\Delta\tilde{\mu}
\end{equation}
per 2D carrier, causing the cyclotron orbits to drift orthogonal to this force. The resulting current is
\begin{equation}\label{inducedcurrent}
j_{y,{\rm m}}=\frac{N_{\rm 2D}}{qB_0}\bigg(\frac{2\eta}{\gamma\beta}\frac{\partial\Delta\tilde{\mu}}{\partial x}\bigg),
\end{equation}
where $N_{\rm 2D}=\varepsilon_{\rm F}g_{\rm 2D}$ is the number density of 2D carriers. This introduces an anomalous contribution to the Hall effect within the bulk,  such that the modified Hall conductivity becomes
\begin{equation}\label{hallconductivity}
\sigma^\prime_{xy}=\sigma_{xy}(1+\tilde{\chi})^{-1},
\end{equation}
where
\begin{equation}\label{quantizedconductance}
\sigma_{xy}=\frac{2\nu e^2}{h}
\end{equation}
has its usual quantized value. The finite susceptibility in which $\partial B_z/\partial H_z=\mu_0(1+\tilde{\chi})$ effectively results in a back-current $j_{y,{\rm m}}=\tilde{\chi}j_{y,{\rm f}}$ in response to the applied (free) current $j_{y,{\rm f}}$, such that the total current is given by 
\begin{equation}\label{current}
j_y=j_{y,{\rm f}}+j_{y,{\rm m}}=(1+\tilde{\chi})j_{y,{\rm f}}.
\end{equation}
On inverting the conductivity tensor in the usual manner, we obtain
\begin{eqnarray}\label{bulkresistivity}
\rho_{yy}^\prime=\frac{(1+\tilde{\chi})^2\sigma_{xx}}{\sigma^2_{xy}+(1+\tilde{\chi})^2\sigma_{xx}\sigma_{yy}~\nonumber}\\
\rho_{xy}^\prime=\frac{(1+\tilde{\chi})\sigma_{xy}}{\sigma^2_{xy}+(1+\tilde{\chi})^2\sigma_{xx}\sigma_{yy}}.
\end{eqnarray}
Here, we assume that $\sigma_{yx}=-\sigma_{xy}$ and that the DW is in equilibrium prior to the application of a current. Given that $|\tilde{\chi}|\lesssim$~0.1, Equation (\ref{bulkresistivity}) can only explain small changes in the Hall ratio and  resistivity.

The induced current given by Equation (\ref{inducedcurrent}) is not expected to directly modify the net resistance of a uniform bar-like sample, since the confinement of the current to the sample causes its net value to average to zero as shown in Fig.~\ref{currents}(c). $\Delta\tilde{\mu}$ and $M$ cannot be finite beyond the sample surface, giving rise to a surface magnetic current that compensates $j_{y,{\rm m}}$ within the bulk. In accordance with the critical state model, this current can only flow at the maximum critical density
\begin{equation}\label{saturationcurrent}
j_{\rm c}=\frac{N_{\rm 2D}}{qB_0}|{\bf F}_{\rm p, max}|
\end{equation}
that the DW pinning force can sustain, causing a critical region of finite width to propagate into the interior of the sample, instead if being confined to the edge.

\subsection{Inhomogeneous transport}
A non-uniform $\beta({\bf r})$ has the potential to radically transform the magneto transport. If, for example, $\beta({\bf r})$ varies spatially in a random fashion such that $\bar{\beta}v=\int_v\beta({\bf r}){\rm d}x{\rm d}y{\rm d}z$ is the average over the volume $v$, there can be regions of finite size where $\beta({\bf r})\rightarrow$~1 and $\beta({\bf r})\rightarrow$~0 are realized. Regions where $\beta({\bf r})\rightarrow$~0 will contribute most significantly to the total conductance, and these are the same regions where locally $\tilde{\chi}({\bf r})\rightarrow-1$, yielding $\rho_{yy}\rightarrow$~0 and $\rho_{xy}\rightarrow$~0 in Equation (\ref{bulkresistivity}). Hence, we would expect the overall conductance to become filamentary or percolative in nature analogous to an inhomogeneous superconductor~\cite{harrison5,testardi1,maza1,veira1,hsu1,bednorz1}. While the form and distribution of $\beta({\bf r})$ throughout the sample remains unknown, the electrical resistivity in Fig.~\ref{transition} has already been shown to be fitted by the same expression,
\begin{equation}\label{errorfunction}
\rho(T)=\frac{1}{2}\bigg({\rm erf}\bigg[\frac{T-T_{\rm c}}{\Delta T_{\rm c}}\bigg]
+{\rm erf}\bigg[\frac{-T-T_{\rm c}}{\Delta T_{\rm c}}\bigg]+2\bigg)\rho_{\rm n}(T),
\end{equation}
that fits broadened transitions in inhomogeneous superconductors to astonishing accuracy~\cite{harrison5,testardi1,maza1,veira1,hsu1,bednorz1}, albeit with a rather wide transition width $\Delta T_{\rm c}$. Here $T_{\rm c}$ is the midpoint of the transition while $\rho_{\rm n}$ is the extrapolated `normal state' resistivity. 

Given that superconductivity was the original prediction for systems with CDW-like modulations~\cite{harrison5,frohlich1}, it is somewhat ironic that the resistivity can be fitted by a model that normally applies to superconductors. The physical mechanism discussed above is nonetheless very different. Another consequence of percolative or filamentary conductance is that the current will take indirect (i.e. meandering) paths throughout the sample. This gives rise to `current jetting' effects in which the net polarity of the voltage between voltage terminals switches with respect to that of the net current between current terminals during a 4 wire transport measurement, as observed in $\alpha$-(BEDT-TTF)$_2$TlHg(SCN)$_4$~\cite{honold1}.

\subsection{Ballistic transport in a bulk chiral metal}
\begin{figure}
\centering \includegraphics*[scale=0.8,angle=0]{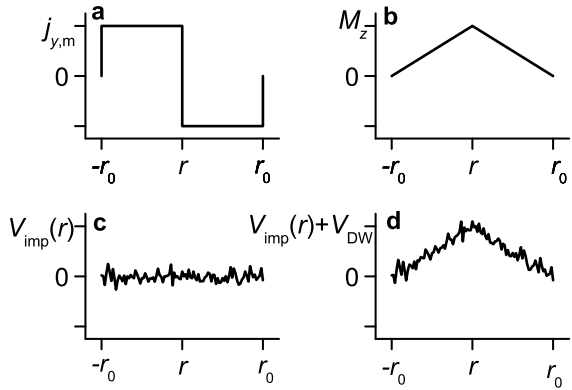}
\caption{Irreversible current (a) and magnetization (b) distributions across a hypothetical cylindrical sample of radius $r_0$ for which the magnetization is fully saturated. The pinning force of the DW introduces an additional effective potential $V_{\rm DW}(r)$ that adds to the impurity potential $V_{\rm imp}({\bf r})$, shown schematically in (c), to arrive at the total potential (d).}
\label{profile}
\end{figure}

Once the critical state is fully established after sweeping the magnetic field over an interval that exceeds double the characteristic coercion field $2H^\ast$ in Fig.~\ref{hysteresisexamples}b, the profile of $j_{y,{\rm m}}$ and the magnetization $M$ will develop the form depicted in Fig.~\ref{profile}. A bulk chiral metal may occur if the net difference in the pinning potential $\Delta V_{\rm DW}=V_{\rm DW}(r_0)-V_{\rm DW}(0)$ (i.e. see Fig.~\ref{profile}) exceeds the overall width of the random impurity potential $V_{\rm imp}({\bf r})$. In such a situation, all of the cyclotron orbits will drift continuously about the sample axis ($r=$~0), giving rise to the possibility of ballistic transport when backscattering becomes inhibited~\cite{balents1}.

The average width of the impurity potential is usually given by $\Delta V_{\rm imp}=\hbar/2\tau$, although the actual spatial variation is system dependent. In $\alpha$-(BEDT-TTF)$_2$KHg(SCN)$_4$, where $\tau^{-1}\approx$~0.2~$\times$~10$^{12}$~s$^{-1}$, this corresponds to $\Delta V_{\rm imp}\approx$~0.07~meV~\cite{harrison8}. $\Delta V_{\rm DW}$ for the situation, depicted in Fig.~\ref{profile}d, can be estimated from the experimental data by using the approximate formula
\begin{equation}\label{DWpotential}
\Delta V_{\rm DW}\approx r_0|{\bf F}_{\rm p}|=\frac{j_{\rm c}r_0B_0}{N_{\rm 2D}}\approx\frac{3M_{\rm sat}B_0}{N_{\rm 2D}},
\end{equation}
where $M_{\rm sat}$ is the saturation magnetization averaged over the volume (assuming a cylindrical approximation).
This yields  $\Delta V_{\rm DW}$ values ranging from $\approx$~1.03~meV for $M_{\rm sat}\approx$~300~Am$^{-1}$in Reference~\cite{harrison3} to $\approx$~0.15~meV for $M_{\rm sat}\approx$~50~Am$^{-1}$ in Reference~\cite{harrison2}. The former estimate exceeds that of $\Delta V_{\rm imp}$ by a factor of $\approx$~15, implying that Fig.~\ref{profile}d provides a realistic model.

Should ballistic transport become a realistic factor in orbitally quantized DW phases~\cite{harrison2,hill2,honold2}, screening of $\Delta B_z$ by currents near the surface of a sample will prevent an energetically costly change in $\Delta\mu$ from developing within the bulk. Setting $\Delta\mu=$~0 in Equation (\ref{nonequilibrium1}) then yields
\begin{equation}\label{penetration}
\lambda^2\nabla^2V-V=0~~~{\rm with}~~~\lambda=[m/(2\mu_0e^2N_{\rm 2D})]^{1/2},
\end{equation}
where ${\bf E}$=$\nabla V$, while $\lambda$ is a penetration depth of similar form to that in superconductors. One consequence of ballistic transport is that the Hall effect becomes `ideal,' so that $\Delta B_z=-\mu_0eN_{\rm 2D}V/B_0$. On estimating $\lambda$ for $\alpha$-(BEDT-TTF)$_2$KHg(SCN)$_4$ we obtain $\lambda\approx$~400~nm~\cite{harrison2}. This quantity is too small to have been measured directly, but has been proposed as an explanation for inductive currents in a $\sim$~1~mm$^2$ cross-section sample with a decay time exceeding 10~$\mu$s  that are heavily weighted towards its edge~\cite{harrison2}.

The spatial translation of the orbitally quantized DW with respect to the edge in the critical state region in Fig.~\ref{profile} as $B_z$ is swept could be another factor in inducing surface-weighted currents. As consecutive wave fronts of the DW slide past the edge they carry with them magnetic flux, causing currents that were previously pinned to become free near the surface. Loss of pinning should lead to the spontaneous generation of a Hall voltage (or redistribution of charge near the surface) to counter the Lorentz force. A similar spontaneous Hall voltage should occur if the critical state collapses due to an increase in the temperature. Indeed, $j_{\rm c}$ is observed to be strongly temperature dependent~\cite{harrison3}. This would provide a new mechanism for pyroelectric currents, that are presently the realm of ferroelectric materials~\cite{lijima1}.

\section{Future directions}
Orbitally quantized DW systems have become widely known for producing a mechanism by which one can observe a variant of the quantum Hall effect in a bulk material. While this is firmly established experimentally in TMTSF-based salts~\cite{hannahs1,chaikin1}, large Landau filling factors of $\nu\gtrsim$~20 have proven to be something of an impediment in observing a quantized Hall resistance in $\alpha$-(BEDT-TTF)$_2M$Hg(SCN)$_4$ salts~\cite{honold1}. Meanwhile, the involvement of orbital quantization in (Per)$_2$Pt(mnt)$_2$~\cite{graf1} and in $\alpha$-(BEDT-TTF)$_2M$Hg(SCN)$_4$ when ${\bf H}$  is aligned within the layers~\cite{andres1} remains to be established.

In the present review we identify another intrinsic property of orbitally quantized DW systems, which is their ability to store energy inductively in an analogous manner to the vortex state, albeit in the presence of a large static background magnetic field. Such effects are prevalent in layered organic conductors because of their narrow electronic bandwidths, causing them to be somewhat easier to manipulate with a magnetic field than conventional CDW and SDW materials~\cite{gruner1}. The magnitude of these effects are nevertheless rather small compared to those in the vortex state, with dissipationless conduction requiring considerable improvements in sample quality. Sample-dependence already appears to be a major factor. Although the in-plane resistivity of some TMTSF-based salts shows a significant drop at low temperatures~\cite{chaikin1}, as might be expected to accompany the quantum Hall effect, this is not universally observed throughout the literature. The same is also true with $\alpha$-(BEDT-TTF)$_2M$Hg(SCN)$_4$. Samples grown by Tokumoto {\it et al}~\cite{harrison3,harrison5} consistently exhibit the most remarkable drops in resistivity on entry into the high magnetic field DW phase, by as much as a factor of 100, closely followed by those of Kurmoo {\it et al.}~\cite{honold2}. The present variability of such effects suggests that there is considerable room for sample improvement, potentially yielding exotic new physical effects. On the other hand, samples that do not exhibit a significant drop in resistivity would be expected to support larger electric effects that would ultimately yield a more conventional non-linear DW conduction to be observed~\cite{fujita1}.

This work is supported by US Department of Energy, the National Science Foundation and the
State of Florida.

\end{document}